PAPER • OPEN ACCESS

# Cloud shape of a molecular Bose–Einstein condensate in a disordered trap: a case study of the dirty boson problem

To cite this article: Benjamin Nagler *et al* 2020 *New J. Phys.* **22** 033021

View the article online for updates and enhancements.

# New Journal of Physics

The open access journal at the forefront of physics

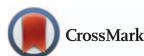

**PAPER**

**OPEN ACCESS**

# Cloud shape of a molecular Bose–Einstein condensate in a disordered trap: a case study of the dirty boson problem

Benjamin Nagler[1,2], Milan Radonjić[1,3], Sian Barbosa[1], Jennifer Koch[1], Axel Pelster[1] and Artur Widera[1,2]

[1] Department of Physics and Research Center OPTIMAS, Technische Universität Kaiserslautern, D-67663 Kaiserslautern, Germany
[2] Graduate School Materials Science in Mainz, Gottlieb-Daimler-Straße 47, D-67663 Kaiserslautern, Germany
[3] Scientific Computing Laboratory, Center for the Study of Complex Systems, Institute of Physics Belgrade, University of Belgrade, Serbia

E-mail: widera@physik.uni-kl.de

## Abstract

We investigate, both experimentally and theoretically, the static geometric properties of a harmonically trapped Bose–Einstein condensate of $^6$Li$_2$ molecules in laser speckle potentials. Experimentally, we measure the *in situ* column density profiles and the corresponding transverse cloud widths over many laser speckle realizations. We compare the measured widths with a theory that is non-perturbative with respect to the disorder and includes quantum fluctuations. Importantly, for small disorder strengths we find quantitative agreement with the perturbative approach of Huang and Meng, which is based on Bogoliubov theory. For strong disorder our theory perfectly reproduces the geometric mean of the measured transverse widths. However, we also observe a systematic deviation of the individual measured widths from the theoretically predicted ones. In fact, the measured cloud aspect ratio monotonously decreases with increasing disorder strength, while the theory yields a constant ratio. We attribute this discrepancy to the utilized local density approximation, whose possible failure for strong disorder suggests a potential future improvement.

## 1. Introduction

Disorder is ubiquitous in nature, since there is no perfect order in any realistic physical system. Random disordered media have for long time been studied in terms of percolation and transport phenomena [1–3] or anomalous diffusion [4]. The dirty boson problem first arose experimentally in the context of superfluid helium in porous vycor glass, where it was shown that superfluidity can still persist despite the influence of disorder [5, 6]. It amounts to understanding the emergence of coherence and order for ultracold bosonic atoms in the presence of frozen disorder in the form of a static random potential. Its importance and intriguing aspects stem from the interplay of localization and interactions as well as of disorder and superfluidity. Since the realization of Bose–Einstein condensates (BECs) in 1995 [7, 8] the interest regarding the effects of disorder potentials on the properties of ultracold quantum gases has substantially increased [9]. This was initiated by unavoidable irregularities in the trapping potential induced by wire imperfections [10, 11]. However, ultracold quantum gases became controllable to an unprecedented level of precision. For example, disorder can nowadays be imposed thereon in a highly tunable manner via laser speckles [12, 13], light fields engineered by digital micromirror devices [14, 15] or atomic impurities trapped in an optical lattice [16, 17]. Hence, they represent an extremely promising platform for practically realizing Feynman's quantum simulator [18], which mimics the physics of other quantum many-body systems [19, 20]. This especially applies to many phenomena of condensed matter physics, which are inevitably influenced, or even caused, by various types of uncontrollable disorder. For instance, the phenomenon of Anderson localization, originally used to microscopically describe the absence of diffusion in the presence of disorder [21], was predicted for a flow of a dilute Bose gas through a disordered region [22]. So far it was observed for non-interacting atoms in one dimension [23, 24] as well as presumably in three dimensions [25, 26]. More recently, the so called many-body localization has been observed for interacting particles in disordered lattice systems and has attracted significant interest [27–29]. In this paper, however, we are more interested in analyzing the impact of disorder upon bosonic many-body quantum systems in the bulk. Therefore, we briefly review the literature on the homogeneous or trapped dirty boson problem both for a weak and a strong random potential.

In order to quantitatively study a weakly interacting homogeneous Bose gas in a static random potential, Huang and Meng worked out a Bogoliubov theory [30], where quantum and thermal fluctuations as well as disorder fluctuations were assumed to be small. This approach treated the disorder perturbatively and was subsequently extended by others, either within the original framework of second quantization [31–34] or within the functional integral approach including the replica method [35–37]. In particular, it was also demonstrated that, despite the randomness of the potential, superfluidity persists and that its depletion is larger than the condensate depletion since the bosons scattered by the disorder landscape represent randomly distributed obstacles for the motion of the superfluid. An extension to the situation when the disorder correlation function falls off with a characteristic correlation length, as in the case of a Gaussian [32, 38–40], a Lorentzian [41], or laser speckle disorder [42, 43] revealed that condensate and superfluid depletions decrease with increasing correlation length. Note that, to the best of our knowledge, so far no prediction of the Huang–Meng theory has yet been checked experimentally although it stems from 1992. Finally, some subtleties regarding condensate deformation and its quantum depletion in external, and possibly random, weak potentials were resolved in [44, 45] using an inhomogeneous Bogoliubov theory.

The dirty boson problem was also studied non-perturbatively using various approaches. For a delta-correlated disorder, either in the homogeneous case [46–51] or in a disordered trap [52–54], a paradigmatic result is that an increase of the disorder strength at zero temperature leads to a first-order quantum phase transition from a superfluid to a Bose-glass phase. In the latter, global phase coherence is destroyed and phase coherence only persists locally due to the particle scattering with the disorder. On the other hand, non-perturbative studies of dirty bosons with a finite disorder correlation length are scarce. For instance, the diffusion Monte-Carlo study in [55], where the disorder is implemented within a hard-sphere model, concludes for the homogeneous case that no quantum phase transition from a superfluid to a Bose-glass occurs. This result is confirmed by the theoretical investigations of this paper, where the case of a disordered trap is studied for concrete experimental parameters.

Note that, until now, it has remained a challenge to test all these theoretical predictions within a concrete experimental setup. This motivated us to investigate within an experiment-theory collaboration the static cloud shape of a BEC in a disordered trap and its change with increasing disorder strength. In section 2 we introduce the underlying experimental setup, where the disorder is realized by laser speckles and where the harmonically trapped BEC consists of molecules of fermionic $^6$Li atoms. Subsequent section 3 develops a corresponding theoretical description, which is non-perturbative with respect to disorder and includes quantum fluctuations. Finally, section 4 presents the measured cloud widths as a function of the laser speckle strength and discusses how they can be explained both qualitatively and quantitatively by the developed theory. Although we have achieved a quite significant level of agreement between the measurements and the theory, we also note that some experimental results indicate certain limitations of our theoretical approach, which needs to be refined for future studies.

## 2. Experimental setup

In the following, relevant components of the experimental setup are presented. First, we give an overview of the sequence for the creation of molecular BECs. Subsequently we focus on statistical properties of the optical speckle potential, its interaction with lithium molecules and a novel method for calibrating the disorder strength. Finally, we describe how to load the quantum gas into a disordered trap, the measurement protocol as well as the data analysis.

### 2.1. Creation of molecular Bose–Einstein condensates

A general overview of our experimental setup is given in [56]. In short, we prepare quantum degenerate gases of neutral, fermionic $^6$Li atoms in an equal mixture of the two lowest-lying Zeeman substates $|m_J = -1/2,\ m_I = 1\rangle$ and $|-1/2, 0\rangle$ of the electronic ground state $^2S_{1/2}$, where $m_J$ and $m_I$ are the magnetic quantum numbers of the electronic and nuclear spin, respectively. A magnetic Feshbach resonance centered at 832 G [57] is utilized to tune the s-wave scattering length between atoms of opposite spin to 4510 $a_0$ at 763.6 G, where $a_0$ denotes the Bohr radius. At this magnetic field, the interatomic interaction potential features a molecular state with binding energy $E_b/k_B = 1.5\ \mu$K [58], where $k_B$ denotes the Boltzmann constant. Such Feshbach molecules form during cooling once the sample temperature approaches $E_b/k_B$ [59] and they interact with scattering length 2706 $a_0$ [60]. Molecular BECs of typically $(4.3 \pm 0.2) \times 10^5$ molecules are created by forced evaporative cooling in a superposition of optical and magnetic potentials. An optical dipole trap confines the cloud radially, which corresponds to the $x$- and $z$-direction in figure 1, with trap frequencies $(\omega_x, \omega_z) = 2\pi \times (195\ \text{Hz}, 129\ \text{Hz})$. The weaker axial confinement, along the $y$-direction in figure 1, is provided by a magnetic saddle potential for which we have $\omega_y = 2\pi \times 22.6\ \text{Hz}$. After evaporation, the atomic sample is held for 500 ms at a constant trap depth of 500 nK $\times\ k_B$ to ensure thermal equilibration, before studying its properties in a disordered potential.

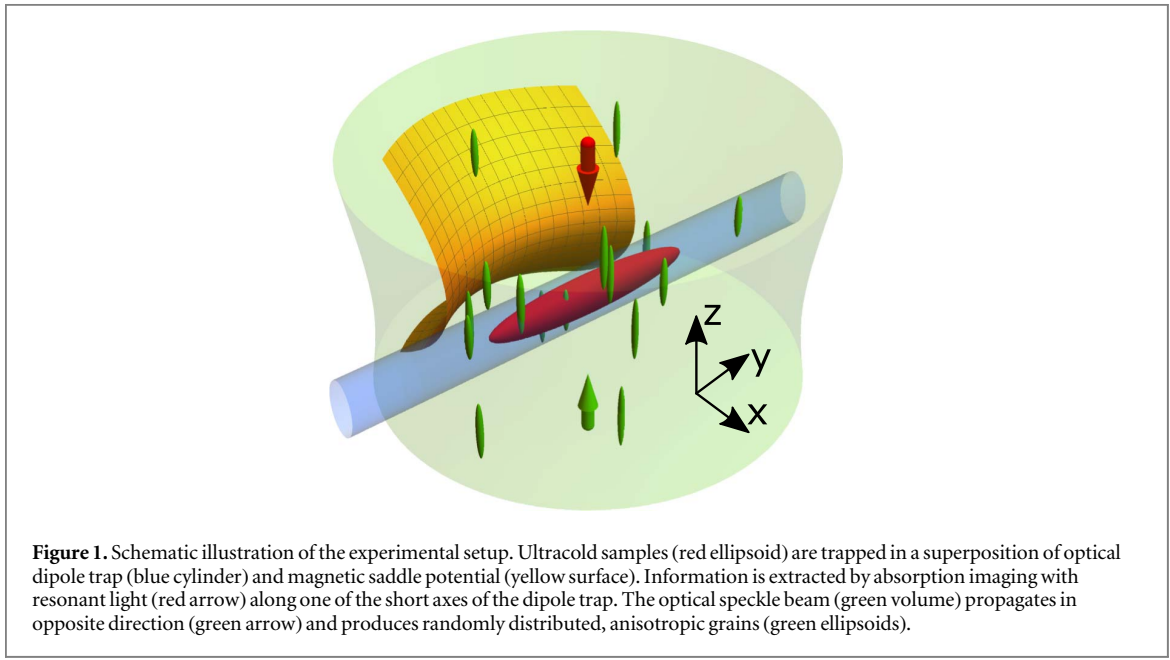

**Figure 1.** Schematic illustration of the experimental setup. Ultracold samples (red ellipsoid) are trapped in a superposition of optical dipole trap (blue cylinder) and magnetic saddle potential (yellow surface). Information is extracted by absorption imaging with resonant light (red arrow) along one of the short axes of the dipole trap. The optical speckle beam (green volume) propagates in opposite direction (green arrow) and produces randomly distributed, anisotropic grains (green ellipsoids).

## 2.2. Optical speckle potential

The optical speckle potential is realized by passing a Gaussian laser beam through a holographic diffusive plate (Edmund Optics 47-991) and imaging the light onto the position of the atoms. The beam is derived from a frequency-doubled Nd:YVO solid-state laser (Coherent Verdi V18) with wavelength $\lambda = 532$ nm. We control its power irradiated onto the atomic sample by means of an acousto-optic modulator in a proportional-integral-derivative control loop. After passing through the diffuser, the beam waist is enlarged to $w = 29$ mm by a telescope to match the free aperture of the imaging objective with numerical aperture 0.29 and focal length $f = 98.8$ mm.

The speckle grain size or correlation length is characterized by the width of the intensity autocorrelation function $C(\delta \mathbf{r}) = \overline{I(\mathbf{r})I(\mathbf{r} + \delta \mathbf{r})}$ [12], where $I(\mathbf{r})$ is the intensity at the point $\mathbf{r}$ and the bar denotes averaging over a region that encompasses a large number of speckle grains. For a Gaussian beam with waist $w$, the autocorrelation function in the image plane takes on the shape of a Gaussian

$$C(\delta \mathbf{r}) = \overline{I}^2 \left[1 + \exp\left(-\frac{\delta \mathbf{r}^2}{\sigma^2}\right)\right], \qquad (1)$$

with the diffraction-limited width $\sigma = \lambda f/(\pi w) = 576$ nm for $w = 29$ mm. Along the beam propagation axis the grain size is expected to be larger by a factor of $4f/w = 13.6$ [61]. A speckle is called fully developed if the phases of contributing partial waves are distributed uniformly across the interval $[0, 2\pi)$ [12]. In this case, dark regions are more abundant than speckle grains of any higher intensity [62] and one can show that the probability density of the intensity follows an exponential law [12]

$$P(I) = \frac{1}{\overline{I}} \exp\left(-\frac{I}{\overline{I}}\right). \qquad (2)$$

Both the intensity autocorrelation and probability distribution are measured in a test setup featuring an optical beam path for projection of the laser speckle that is identical to the one used in the experiment. It includes a microscope at the intended atomic cloud position which images the speckle onto a CCD camera. The measured intensity probability distribution, shown in figure 2, is exponential. Thus, we conclude that the speckle is, indeed, fully developed. The measured correlation length $\sigma_m = 750$ nm is larger than theoretically expected, which is in part due to a truncation of the Gaussian beam by the finite aperture of the objective. Taking this into account, a numerical simulation of the speckle provides a correlation length of 650 nm. The remaining deviation is attributed to potential aberrations of our optical system.

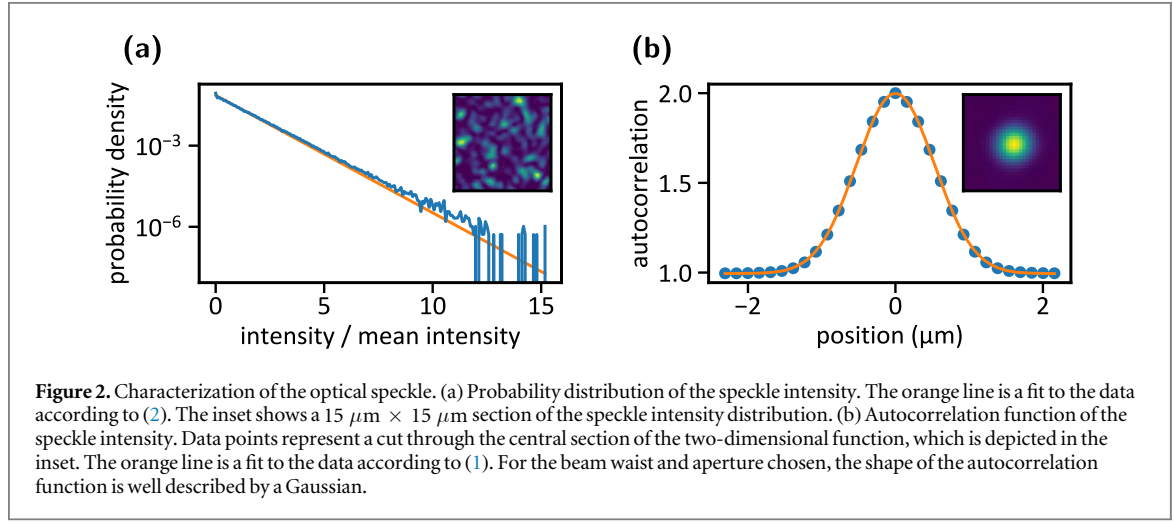

**Figure 2.** Characterization of the optical speckle. (a) Probability distribution of the speckle intensity. The orange line is a fit to the data according to (2). The inset shows a 15 $\mu$m × 15 $\mu$m section of the speckle intensity distribution. (b) Autocorrelation function of the speckle intensity. Data points represent a cut through the central section of the two-dimensional function, which is depicted in the inset. The orange line is a fit to the data according to (1). For the beam waist and aperture chosen, the shape of the autocorrelation function is well described by a Gaussian.

Since the atomic transition of lithium, which is affected the most by the laser speckle, is the far detuned D2 line at 671 nm, the interaction with the light field is dominated by the optical dipole force [63]. The corresponding potential reads

$$u(\mathbf{r}) = -\frac{3\pi c^2 \Gamma}{2\omega_0^3}\left(\frac{1}{\omega_0 - \omega} + \frac{1}{\omega_0 + \omega}\right)I(\mathbf{r}), \qquad (3)$$

with the speed of light $c$, the natural line width $\Gamma = 2\pi \times 5.87$ MHz [64], the speckle laser frequency $\omega = 2\pi c/\lambda$, and the atomic transition frequency $\omega_0$. As $\omega > \omega_0$, the potential is repulsive and we define the disorder strength as the mean potential $\bar{u}$ induced by the mean intensity $\bar{I}$. Spontaneous scattering is negligible ($<10^{-3}$ s$^{-1}$) for all intensities used in our experiments. The polarisability of $^6$Li$_2$ molecules is twice that of unbound atoms [65] and, therefore, the molecules experience double the dipole potential. The maximal disorder strength for molecules we can achieve with the available laser power is $\bar{u}/k_B = 356$ nK.

Another relevant characteristic of optical speckle is the spatial dependence of its mean intensity $\bar{I}$. Due to the finite scattering angle $\vartheta$ of the diffusive plate, the light field in the image plane extends over a range far greater than the diffraction-limited spot size and its envelope is described by a Gaussian beam shape with waist $w_s = \vartheta f$ [12]. In our setup, we have $\vartheta = 0.5°$ and therefore we obtain $w_s = 862$ $\mu$m, which is larger than the typical size 300 $\mu$m of an ultracold atomic cloud. For a precise calibration of the mean intensity it is therefore crucial not only to know the speckle waist but also the cloud's position therein. Other calibration methods are restricted to attractive potentials [66] or rely on measurements of the momentum distribution [67] and are, thus, not applicable to our system.

We have developed a novel method, which enables us to align the speckle center to the atomic cloud and infer the waist simultaneously. It is based on rotating the speckle pattern as a whole and measuring the resulting displacement of an atomic cloud within the speckle, see figure 3. Our diffusive plate is attached to a motorized rotary stage (Thorlabs K10CR1/M). Spinning this stage causes the emerging speckle pattern to revolve around its principal axis in the image plane. For the alignment procedure, we create a molecular BEC and linearly ramp the speckle laser power to its maximum value during 50 ms. The sample is held for 100 ms in the potential rotating with angular velocity $\omega = 20°$ s$^{-1}$ before being released from the trap and imaged along the speckle principal axis after 11 ms time of flight. Since the disorder strength around the envelope center at $x = y = 0$ is large enough to drag the cloud along with its motion, the displacement of the atoms $\Delta s$ is approximately the distance traversed by the speckle $\Delta d(x, y) = \omega\sqrt{x^2 + y^2} t_i$ during the illumination time $t_i$. The value $t_i = 136$ ms takes into account the weighted ramp duration and the time of flight. For increasing distance from the center, the decrease in mean intensity will eventually lead to a decline of $\Delta s$, although $\Delta d$ continues to grow. We model this situation by assuming $\Delta s$ to be, up to a constant factor, given by the product of $\Delta d$ and the envelope function:

$$\Delta s(x, y) \propto \Delta d(x, y) \frac{\bar{I}(x, y)}{\bar{I}(0, 0)} = \Delta d(x, y) \exp\left(-2\frac{x^2 + y^2}{w_s^2}\right). \qquad (4)$$

In order to infer $w_s$, we scan the relative position of the quantum gas with respect to the speckle potential across the $x$- or $y$-axis. For each position, we measure $\Delta y$ or $\Delta x$ for clockwise ($\circlearrowright$) and anticlockwise ($\circlearrowleft$) rotation and evaluate $\Delta y_{\text{diff}} = \Delta y_\circlearrowright - \Delta y_\circlearrowleft$, where $\Delta y_\circlearrowright \approx -\Delta y_\circlearrowleft$, or $\Delta x_{\text{diff}}$ in order to increase the signal strength. The result of such a measurement is shown in figures 3(b) and (c), where each data point is the average over 10 repetitions for each rotational direction. From fits according to (4) along the respective axes we extract $w_{s,x} = 808(38)$ $\mu$m and $w_{s,y} = 884(63)$ $\mu$m, which agrees with $w_s = 862$ $\mu$m within less than 10% deviation. Note that the smaller displacement along *y* compared to *x* is due to an additional confining potential that is not turned off during time of flight. As a possible cause for the ellipticity we identify a slight astigmatism induced by the acousto-optic modulator. For further experiments, we shift the peak of the speckle envelope to the atomic cloud position in order to achieve an almost homogeneous random potential with about 6% variation of $\bar{I}$ across the typical cloud size.

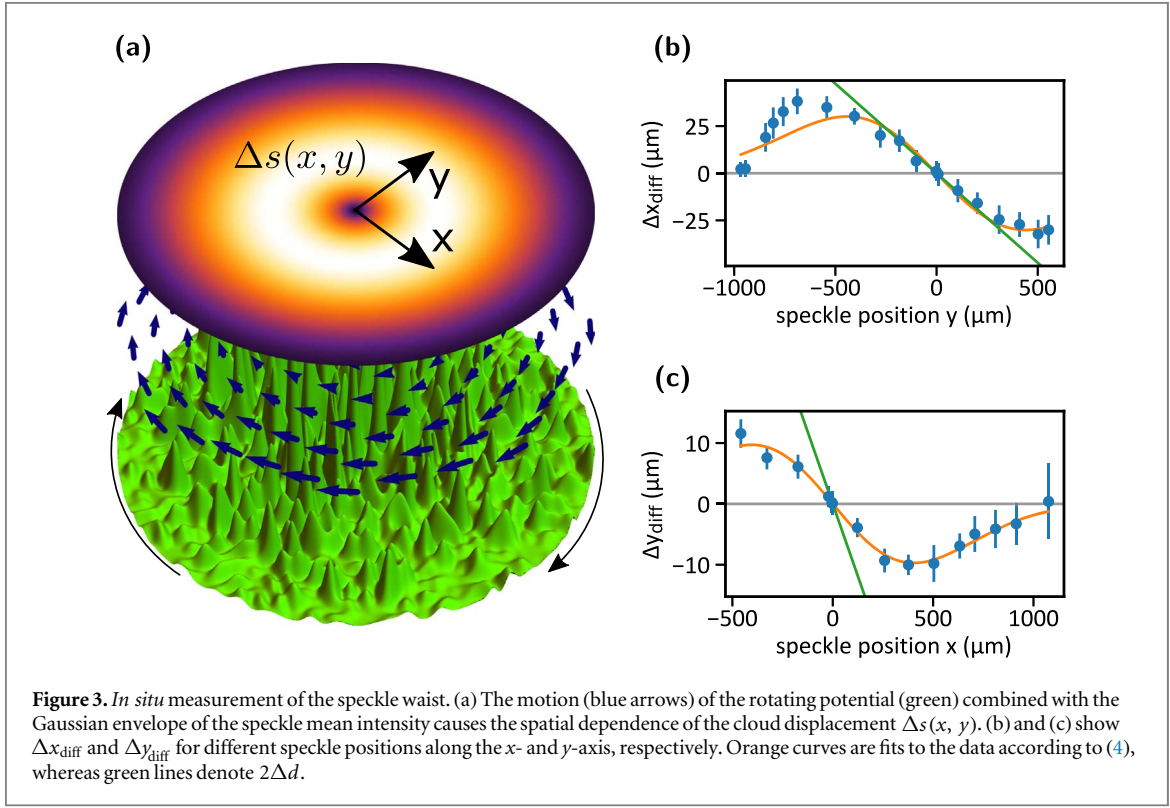

**Figure 3.** *In situ* measurement of the speckle waist. (a) The motion (blue arrows) of the rotating potential (green) combined with the Gaussian envelope of the speckle mean intensity causes the spatial dependence of the cloud displacement $\Delta s(x, y)$. (b) and (c) show $\Delta x_{\text{diff}}$ and $\Delta y_{\text{diff}}$ for different speckle positions along the *x*- and *y*-axis, respectively. Orange curves are fits to the data according to (4), whereas green lines denote $2\Delta d$.

### 2.3. Molecular Bose–Einstein condensates in a disordered trap

In order to load the molecular BEC into the disordered potential, the speckle potential is linearly ramped to its final value $\bar{u}/k_B$ within the interval [0, 356] nK during 50 ms. After an additional hold time of 100 ms we probe the system by performing resonant high-intensity absorption imaging [68] on the transition $|m_J = -1/2, m_I = 1\rangle \leftrightarrow |^2P_{3/2}, m_J = -3/2, m_I = 1\rangle$ along the *z*-axis and, therefore, measuring the *in situ* atomic column density in the *x*–*y* plane. We repeat this measurement about 140 times for each sampled disorder strength. Before every iteration, the microscopic details of the random potential experienced by the cloud are altered by rotating the speckle plate. This procedure enables us to average observables over many disorder realizations and, therefore, conform with the theoretical model as described in the subsequent section 3. From a semi-classical fit [69] to integrated density profiles without disorder we estimate the temperature to be 70 nK, which amounts to about 25% of the non-interacting critical temperature for the onset of trapped molecular BEC. As this fit does not account for the quantum depletion of the BEC, the determined temperature has to be considered as an upper bound.

Figure 4 depicts exemplary density profiles for different disorder strengths. The impact of increasing disorder manifests itself in the form of emerging density fluctuations and an alteration of the cloud shape, which is characterized by the spatial extension along the principal axes. Since the healing length $\xi = 270$ nm of the BEC in the trap center is smaller than all correlation lengths, the wave function resolves the finest structures of the speckle potential [70]. These are, however, not visible in the recorded density distributions due to the limited resolution of our imaging setup, which amounts to 2.2 $\mu$m. In order to extract the widths and simultaneously reduce the noise introduced by the fluctuations, we integrate the 2D column density profiles from figure 4(a) along the *y*-direction and fit a Gaussian function from which we get the $1/e$ half-widths $a_x$. Analogously, we obtain $a_y$ from Gaussian fits of the density profiles integrated along the *x*-direction. Although the interaction strength is quite large, this ansatz is well suited to describe the integrated profiles even for the largest explored disorder strength, as illustrated in figures 4(b) and (c).

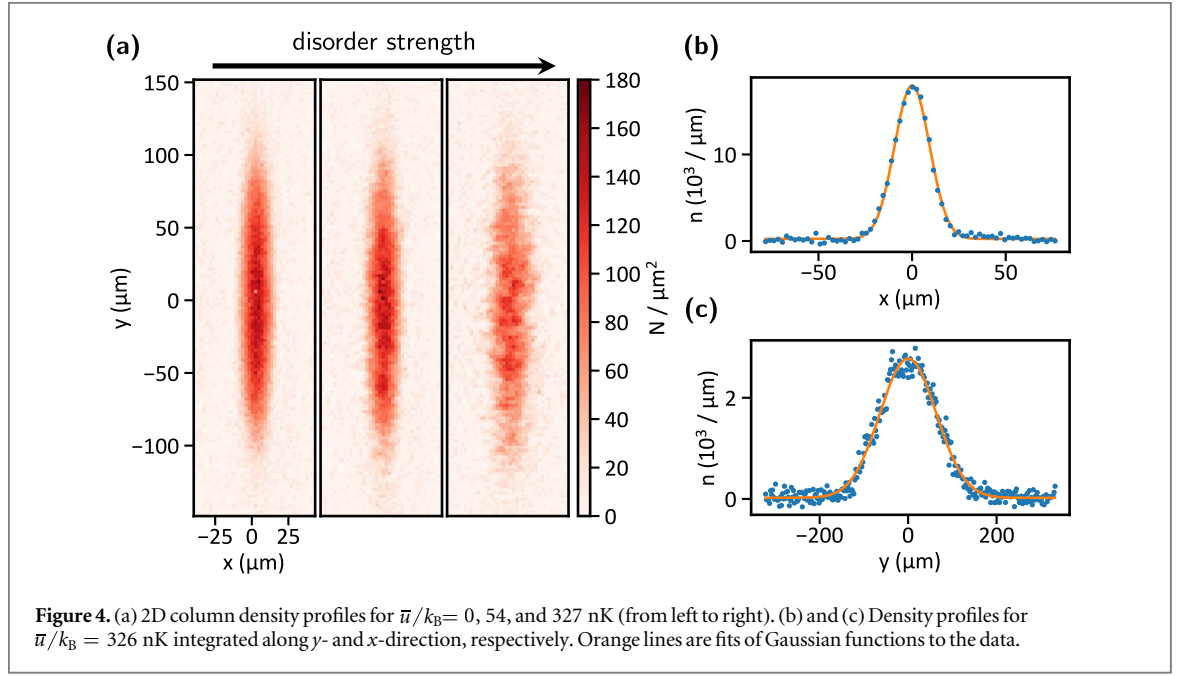

**Figure 4.** (a) 2D column density profiles for $\bar{u}/k_B =$ 0, 54, and 327 nK (from left to right). (b) and (c) Density profiles for $\bar{u}/k_B = 326$ nK integrated along $y$- and $x$-direction, respectively. Orange lines are fits of Gaussian functions to the data.

## 3. Theory

Let us consider a weakly interacting Bose gas at $T = 0$ in the ground state in an external trapping potential $U(\mathbf{r})$ with the addition of a disorder potential $u(\mathbf{r})$. We assume that the disorder is homogeneous after the ensemble average, i.e. $\langle u(\mathbf{r}) \rangle = \langle u \rangle$. Otherwise, any non-homogeneity of the average can be absorbed into $U(\mathbf{r})$. Since the correlation length is much smaller than the spatial extension of the cloud in $x$- and $y$-direction, we conclude that $\langle u \rangle$ matches the spatial average $\bar{u}$. The maximal value of the gas parameter in our experimental setup is $n(\mathbf{0})a_s^3 = 0.017$, where $n(\mathbf{0})$ denotes the clean-case particle density in the trap center, see (6) below. Comparing with the typical gas parameters for weakly interacting $^{87}$Rb gas of about 0.0007 [71] and for a strongly interacting $^4$He of about 0.24 [72], we conclude that our experiment is in an intermediate regime, where quantum fluctuations have to be accounted for. Hence, the ground state wave function $\psi(\mathbf{r})$ satisfies the extended Gross–Pitaevskii equation that includes the quantum-fluctuation corrections of Lee et al [73] on the mean-field level [74]

$$-\frac{\hbar^2}{2m}\nabla^2\psi(\mathbf{r}) + U(\mathbf{r})\psi(\mathbf{r}) + u(\mathbf{r})\psi(\mathbf{r}) - \mu\psi(\mathbf{r}) + g\left[\psi(\mathbf{r})^3 + \frac{40}{3\sqrt{\pi}}a_s^{3/2}\psi(\mathbf{r})^4\right] = 0, \quad (5)$$

where $a_s$ is the $s$-wave scattering length and $g = 4\pi\hbar^2 a_s/m$ denotes the interaction strength. The ground state chemical potential $\mu$ is taken to be independent of the disorder realization. This leads to a distribution of the total number of particles for different realizations of the disorder. The value of $\mu$ is fixed by the requirement that on average one has

$$N = \int_{\mathbb{R}^3} d^3\mathbf{r}\, \langle n(\mathbf{r}) \rangle, \quad \text{with} \quad n(\mathbf{r}) = \psi(\mathbf{r})^2 + \frac{8}{3\sqrt{\pi}}a_s^{3/2}\psi(\mathbf{r})^3, \quad (6)$$

where the second term of the particle density $n(\mathbf{r})$ accounts for the quantum depletion of the condensate [74].

Averaging the condensate wave function over many disorder realizations gives $\langle \psi(\mathbf{r}) \rangle$ that satisfies the average of (5)

$$-\frac{\hbar^2}{2m}\nabla^2\langle\psi(\mathbf{r})\rangle + U(\mathbf{r})\langle\psi(\mathbf{r})\rangle + \langle u(\mathbf{r})\psi(\mathbf{r})\rangle - \mu\langle\psi(\mathbf{r})\rangle$$
$$+ g\left[\langle\psi(\mathbf{r})^3\rangle + \frac{40}{3\sqrt{\pi}}a_s^{3/2}\langle\psi(\mathbf{r})^4\rangle\right] = 0. \quad (7)$$

Let us define $\varphi(\mathbf{r}) = \psi(\mathbf{r}) - \langle\psi(\mathbf{r})\rangle$ to be the part of $\psi(\mathbf{r})$ that fluctuates due to disorder. It obeys $\langle\varphi(\mathbf{r})\rangle = 0$, $\langle\varphi(\mathbf{r})^2\rangle = \langle\psi(\mathbf{r})^2\rangle - \langle\psi(\mathbf{r})\rangle^2$ and satisfies

$$-\frac{\hbar^2}{2m}\nabla^2\varphi(\mathbf{r}) + U(\mathbf{r})\varphi(\mathbf{r}) + (u(\mathbf{r})\psi(\mathbf{r}) - \langle u(\mathbf{r})\psi(\mathbf{r})\rangle) - \mu\varphi(\mathbf{r})$$
$$+ g\left[\psi(\mathbf{r})^3 - \langle\psi(\mathbf{r})^3\rangle + \frac{40}{3\sqrt{\pi}}a_s^{3/2}(\psi(\mathbf{r})^4 - \langle\psi(\mathbf{r})^4\rangle)\right] = 0. \quad (8)$$

From the previous two equations one observes that the averages and the fluctuations are mutually coupled. Moreover, $\langle\psi(\mathbf{r})\rangle$ depends on $\langle u(\mathbf{r})\psi(\mathbf{r})\rangle$ and higher-order averages like $\langle\psi(\mathbf{r})^3\rangle$ and $\langle\psi(\mathbf{r})^4\rangle$. The latter signals the necessity to deal with an infinite hierarchy of averages of increasing order. In order to use analytical tools, we will introduce several simplifying approximations.

First, we are going to use the cumulant expansion method [75–77], where in the Gaussian approximation all the cumulants of higher than second order are neglected. The general expansion method and the approximation procedure are outlined in appendix A. In this manner we have, for instance, $\langle\psi(\mathbf{r})^3\rangle \approx 3\langle\psi(\mathbf{r})\rangle\langle\psi(\mathbf{r})^2\rangle - 2\langle\psi(\mathbf{r})\rangle^3$ and $\langle\psi(\mathbf{r})^4\rangle \approx 3\langle\psi(\mathbf{r})^2\rangle^2 - 2\langle\psi(\mathbf{r})\rangle^4$. Such a procedure truncates the hierarchy of the disorder averages of increasing order and enables an approximate but essentially non-perturbative treatment of the disorder up to the second cumulant order. Thus, applying the cumulant expansion to (7) leads to

$$-\frac{\hbar^2}{2m}\nabla^2\langle\psi(\mathbf{r})\rangle + (U(\mathbf{r}) - \mu + 3g\langle\psi(\mathbf{r})^2\rangle - 2g\langle\psi(\mathbf{r})\rangle^2)\langle\psi(\mathbf{r})\rangle$$
$$+ \langle u(\mathbf{r})\psi(\mathbf{r})\rangle + \frac{40g}{3\sqrt{\pi}}a_s^{3/2}(3\langle\psi(\mathbf{r})^2\rangle^2 - 2\langle\psi(\mathbf{r})\rangle^4) = 0, \quad (9)$$

while the normalization condition (6) becomes

$$N = \int_{\mathbb{R}^3} d^3\mathbf{r}\left[\langle\psi(\mathbf{r})^2\rangle + \frac{8}{3\sqrt{\pi}}a_s^{3/2}(3\langle\psi(\mathbf{r})\rangle\langle\psi(\mathbf{r})^2\rangle - 2\langle\psi(\mathbf{r})\rangle^3)\right]. \quad (10)$$

Secondly, we assume that all the disorder-averaged quantities vary in space much more slowly than their fluctuations do. The smaller the correlation length with respect to the other characteristic lengths of the system, the better this assumption is justified. Therefore, in order to determine $\langle u(\mathbf{r})\psi(\mathbf{r})\rangle$ and $\langle\psi(\mathbf{r})^2\rangle$ from (8) we employ the local density approximation (LDA) by treating $\langle\psi(\mathbf{r})\rangle$, $\langle\psi(\mathbf{r})^2\rangle$, etc, alongside with $U(\mathbf{r})$ as constant background quantities $\langle\psi\rangle$, $\langle\psi^2\rangle$, and $U$, respectively, that will at the end of calculation be restored to their original values at the spatial point $\mathbf{r}$. Of course, $\langle\psi(\mathbf{r})\rangle$ and $\langle\psi(\mathbf{r})^2\rangle$ have to be determined in conjunction with (9) and (10). In this way, (8) under LDA describes the fluctuations of a spatially infinite condensate in a fixed self-consistent locally determined background. The final resulting equation, after applying the cumulant expansion as well, is given in appendix B.

Furthermore, according to the experimental data, the correlation length of the disorder potential in the propagation direction is about one order of magnitude larger than that in the transverse speckle plane. Hence, we will consider the scenario of a quasi two-dimensional disorder that is constant along the *z*-axis and has a Gaussian correlation function in the transverse *x*–*y* plane. After a lengthy but straightforward calculation, which is sketched in appendix B, we obtain

$$\langle u(\mathbf{r})\psi(\mathbf{r})\rangle = \langle u\rangle\langle\psi(\mathbf{r})\rangle - \langle\psi(\mathbf{r})\rangle\int_{\mathbb{R}^2}\frac{d^2\mathbf{k}_\perp}{(2\pi)^2}\frac{\langle u, u\rangle_c(\mathbf{k}_\perp)}{\varepsilon_{\mathbf{k}_\perp} + V(\mathbf{r}) + V_Q(\mathbf{r})}, \quad (11)$$

$$\langle\psi(\mathbf{r})^2\rangle = \langle\psi(\mathbf{r})\rangle^2 + \langle\psi(\mathbf{r})\rangle^2\int_{\mathbb{R}^2}\frac{d^2\mathbf{k}_\perp}{(2\pi)^2}\frac{\langle u, u\rangle_c(\mathbf{k}_\perp)}{(\varepsilon_{\mathbf{k}_\perp} + V(\mathbf{r}) + V_Q(\mathbf{r}))^2}, \quad (12)$$

where $\varepsilon_{\mathbf{k}_\perp} = \hbar^2\mathbf{k}_\perp^2/(2m) = \hbar^2(k_x^2 + k_y^2)/(2m)$ denotes the transverse free-particle dispersion, $V(\mathbf{r}) = U(\mathbf{r}) + \langle u\rangle - \mu + 3g\langle\psi(\mathbf{r})^2\rangle$ has the role of an effective potential and $V_Q(\mathbf{r}) = \frac{160g}{3\sqrt{\pi}}a_s^{3/2}\langle\psi(\mathbf{r})\rangle(3\langle\psi(\mathbf{r})^2\rangle - 2\langle\psi(\mathbf{r})\rangle^2)$ is a correction due to quantum fluctuations. The quantity $\langle u, u\rangle_c(\mathbf{k}_\perp)$ is the Fourier transform of the second-order cumulant of the disorder potential $\langle u, u\rangle_c(\mathbf{r}_\perp - \mathbf{r}'_\perp) \equiv \langle u(\mathbf{r}_\perp), u(\mathbf{r}'_\perp)\rangle_c = \langle u(\mathbf{r}_\perp)u(\mathbf{r}'_\perp)\rangle - \langle u\rangle^2$. Had we considered higher-order cumulants, they would have as well contributed to (11) and (12). The set of equations (9)–(12) has to be solved upon specifying the total number of particles as well as the external and the disorder potential.

In accordance with the experimental situation, we assume that the Bose gas is trapped in an external harmonic potential

$$U(\mathbf{r}) = \frac{m}{2}(\omega_x^2 x^2 + \omega_y^2 y^2 + \omega_z^2 z^2), \quad (13)$$

where $\omega_x$, $\omega_y$, and $\omega_z$ are the corresponding trapping frequencies. In addition, we consider the isotropic 2D speckle disorder that has a Gaussian correlation function of the underlying electric field in the transverse plane

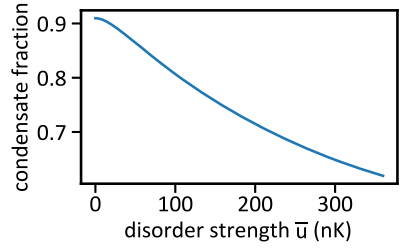

**Figure 5.** Condensate fraction as a function of disorder strength for the experimental parameters.

$$\gamma(\mathbf{r}_\perp) = \frac{\langle \mathcal{E}^*(\mathbf{r}_\perp + \mathbf{r}'_\perp)\mathcal{E}(\mathbf{r}'_\perp)\rangle}{\langle |\mathcal{E}(\mathbf{r}'_\perp)|^2\rangle} = \exp\left(-\frac{\mathbf{r}_\perp^2}{2\sigma^2}\right), \quad (14)$$

where $\sigma$ is the correlation length. Since the resulting speckle potential is proportional to the square of the field, i.e. $u(\mathbf{r}) \propto |\mathcal{E}(\mathbf{r})|^2$, it does not have a Gaussian distribution. However, due to the Gaussian character of the field $\mathcal{E}(\mathbf{r})$, all potential correlation functions can be expressed via sums of products of the field correlation function (14) [78]. Hence, we find

$$\langle u(\mathbf{r}_\perp), u(\mathbf{r}'_\perp)\rangle_c = \langle u\rangle^2 |\gamma(\mathbf{r}_\perp - \mathbf{r}'_\perp)|^2 = \langle u\rangle^2 \exp\left[-\frac{(\mathbf{r}_\perp - \mathbf{r}'_\perp)^2}{\sigma^2}\right], \quad (15)$$

where $\langle u\rangle = \langle u(\mathbf{r}_\perp)\rangle$ is the average speckle potential value. In the transverse $\mathbf{k}_\perp$-plane, we get

$$\langle u, u\rangle_c(\mathbf{k}_\perp) = \pi \sigma^2 \langle u\rangle^2 \exp\left(-\frac{\sigma^2 \mathbf{k}_\perp^2}{4}\right). \quad (16)$$

Now we aim for an approximate semi-analytic variational solution of the coupled set of equations (9)–(12). As no obvious energy functional exists, we cannot rely upon the standard variational procedure of [79, 80], which has been so successful in the realm of ultracold quantum gases. Instead, we follow [81, 82], where a similar variational solution is worked out directly on the basis of the underlying equations of motion within a so-called cumulant approach. To this end, we use a Gaussian ansatz for the disorder-ensemble averaged condensate wave function

$$\langle \psi(\mathbf{r})\rangle = \frac{A}{\pi^{3/4}\sqrt{w_x w_y w_z}} \exp\left(-\frac{x^2}{2w_x^2} - \frac{y^2}{2w_y^2} - \frac{z^2}{2w_z^2}\right), \quad (17)$$

as this reduces the complexity of the problem (9)–(12) to determining only five unknown parameters: the condensate widths $w_x$, $w_y$, and $w_z$, the normalization constant $A$, whose square quantifies the number of bosons in the condensate, and the chemical potential $\mu$. Next, we multiply (9) by $\langle\psi(\mathbf{r})\rangle$, $x^2\langle\psi(\mathbf{r})\rangle$, $y^2\langle\psi(\mathbf{r})\rangle$, and $z^2\langle\psi(\mathbf{r})\rangle$, respectively, and integrate over the whole space. This provides four algebraic equations for the unknowns. The fifth equation is given by the normalization condition (10).

As a first result, we show in figure 5 how the condensate fraction $A^2/N$ monotonously decreases with increasing disorder strength $\bar{u}$, which equals $\langle u\rangle$, for the typical experimental parameters mentioned in section 2. For vanishing disorder we obtain a quantum depletion of 9% for the molecular BEC, which is much larger than for the usual atomic counterparts. Furthermore, in the range of experimentally accessible disorder strengths, i.e. up to $\bar{u}/k_B = 356$ nK, the condensate fraction only drops down to 62%. Even for much larger values of the disorder strength, our theory shows that the condensate fraction never decreases below 45%. With this we conclude that a significant part of the molecular Bose gas remains globally phase coherent despite the mutual impact of both quantum depletion and disorder fluctuations. Thus, as the condensate fraction is far from vanishing, a quantum phase transition from a superfluid to a Bose-glass does not occur for any disorder strength.

In the following we focus our analysis to the cloud widths, which are directly accessible observables. To this end, we first have to calculate the cloud density profile, that is the integrand of (10). Then we find the $1/e$ half-widths along the $x$-, $y$- and $z$-axis of the density profile. In the subsequent section we compare the obtained theoretical cloud widths with the corresponding experimental ones.

## 4. Measurement of the cloud shape in a disordered trap

A comparison of our measurements of the cloud widths along the $x$- and $y$-axis for different disorder strengths with theoretical predictions is presented in figure 6. According to section 3, our theory provides the disorder-ensemble averaged density profiles and their corresponding widths non-perturbatively with respect to disorder, including the quantum fluctuations. On the other hand, in the experiment we determine for each disorder realization the projected density profile and its widths. From the measured width distributions we obtain their average values and standard deviations, respectively. Thus, the experimental and theoretical procedures to obtain the cloud widths are not exactly the same. However, we have carefully checked that an additional analysis of the experimental data in the same manner as in the theoretical model would lead, for all practical purposes, to the same quantitative results for the average widths. Namely, according to the upper plot of figure 6(b), the cloud widths extracted from the averaged density profiles in the range of the considered disorder strengths turn out to deviate only up to 2% from the averages of the individual cloud widths. This deviation is smaller than the statistical error of 5% for extracting the average widths.

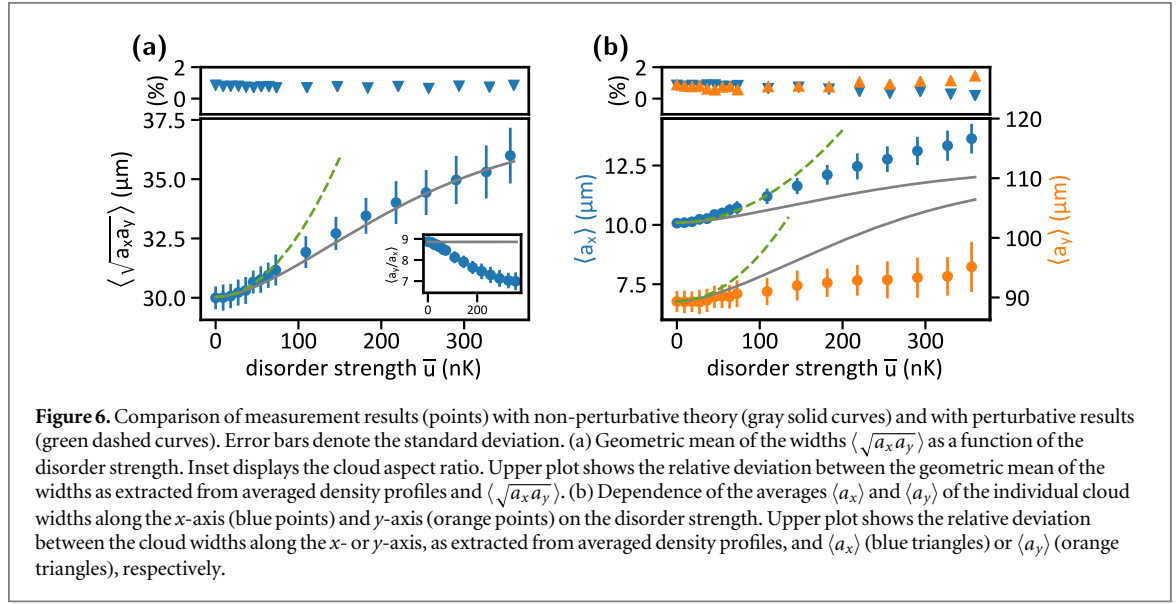

**Figure 6.** Comparison of measurement results (points) with non-perturbative theory (gray solid curves) and with perturbative results (green dashed curves). Error bars denote the standard deviation. (a) Geometric mean of the widths $\langle\sqrt{a_x a_y}\rangle$ as a function of the disorder strength. Inset displays the cloud aspect ratio. Upper plot shows the relative deviation between the geometric mean of the widths as extracted from averaged density profiles and $\langle\sqrt{a_x a_y}\rangle$. (b) Dependence of the averages $\langle a_x\rangle$ and $\langle a_y\rangle$ of the individual cloud widths along the $x$-axis (blue points) and $y$-axis (orange points) on the disorder strength. Upper plot shows the relative deviation between the cloud widths along the $x$- or $y$-axis, as extracted from averaged density profiles, and $\langle a_x\rangle$ (blue triangles) or $\langle a_y\rangle$ (orange triangles), respectively.

In figure 6(a) we find that the geometric mean of the experimentally measured widths $\sqrt{a_x a_y}$, i.e. the cross-section area of the cloud in the imaging plane, is well reproduced by our theory. Additionally, our calculations reveal that the width $a_z$ in the propagation direction also displays qualitatively the same growth. Hence, the cloud steadily expands while the disorder strength is increased. The corresponding particle number of $4.5 \times 10^5$ is determined from the widths in the clean case, i.e. for $\bar{u} = 0$, and is used in the subsequent calculations for all disorder strengths. Its agreement with the experimentally measured value $(4.3 \pm 0.2) \times 10^5$ *a posteriori* justifies the necessity to include quantum fluctuations in our model. Had we neglected quantum fluctuation effects, the theoretically inferred particle number would have been $6.5 \times 10^5$, which is about 45% larger than in the experiment.

In addition, for small disorder strengths the final equations for the unknown parameters $w_x$, $w_y$, $w_z$, $A$, and $\mu$ are solved perturbatively up to the second order in $\bar{u}$. Such an approach could be considered as an extension of the homogeneous dirty boson theory of Huang and Meng [30] to the trapped scenario. We find that the geometric mean of the measured transverse widths is reproduced quite well for $\bar{u}/k_B$ up to about 30 nK, see the green dashed curves in figure 6(a). To the best of our knowledge, this represents the first experimental verification of the Huang–Meng theory from 1992.

Furthermore, we observe in figure 6(b) that the averages of both transverse widths monotonously increase with the disorder strength in the experiment, as well as in the theoretical description. For small enough disorder we find again an agreement with the extended Huang–Meng-like theory. However, we have also to note that the present theory does not reproduce quantitatively the dependence of the individual widths for larger disorder strengths. The smaller width in $x$-direction is systematically underestimated theoretically, while the larger one in $y$-direction is overestimated. For the largest disorder strength the relative differences with respect to the measured values is about 12% for both widths. This noticeable discrepancy is attributed to the fact that the cloud aspect ratio turns out to decrease with increasing disorder strength, while our LDA-based theory predicts a constant aspect ratio, the same as in the clean-case, see the inset of figure 6(a). A physical explanation of this finding is that the disorder potential tends, on average, to make the cloud isotropic as the correlation lengths in both $x$- and $y$-direction coincide. Thus, the transverse isotropic disorder competes with the anisotropic trap and, by increasing its strength, the cloud becomes more and more circular. At the same time, the disorder counteracts the trap confinement and tends to spread out the cloud, which also explains the above mentioned disorder-induced cloud expansion. In our theoretical model, the cloud inevitably inherits the trap anisotropy due to the applied LDA. Hence, in order to capture the measured behavior, one must go beyond the present LDA by, for instance, including systematically the gradients of the confining potential $U(\mathbf{r})$ and of the disorder-ensemble averages $\langle\psi(\mathbf{r})\rangle$ and $\langle\psi(\mathbf{r})^2\rangle$. However, that would make the calculations considerably more numerically demanding.

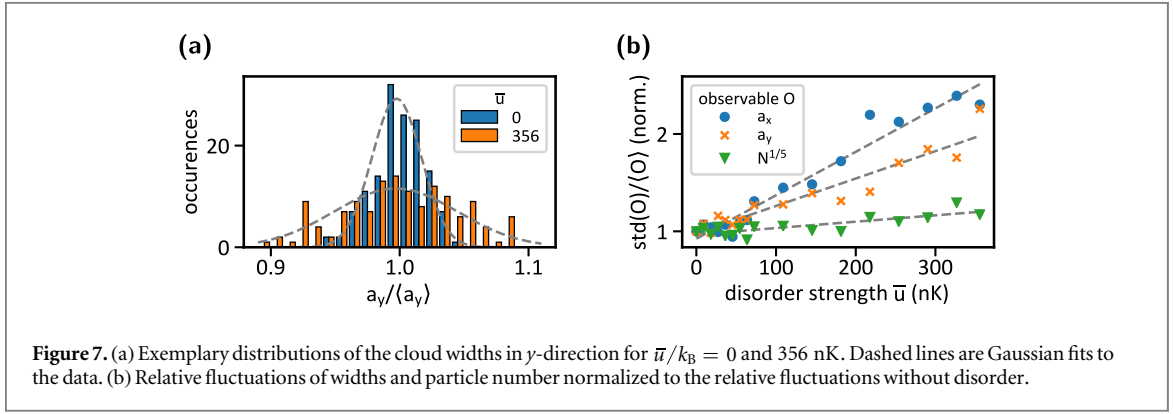

**Figure 7.** (a) Exemplary distributions of the cloud widths in *y*-direction for $\bar{u}/k_B = 0$ and 356 nK. Dashed lines are Gaussian fits to the data. (b) Relative fluctuations of widths and particle number normalized to the relative fluctuations without disorder.

Beyond the so far presented comparison of the data common to both experiment and theory, further experimental measurements (theoretical considerations) can provide additional insights into the disordered Bose gas, that are currently not accessible to theory (experiment). Let us substantiate this conclusion by mentioning exemplarily the following two points.

First, we can easily extract theoretical values for disorder-ensemble averaged condensate widths, which elude any experimental observation as the condensate density is not measurable with the current setup. For weak disorder the condensate widths are closely following the cloud widths. In addition, for the largest experimentally accessible disorder strength $\bar{u}/k_B = 356$ nK the condensate widths turn out to be still 91% of the cloud widths in both *x*- and *y*-direction. This finding is in line with the result from figure 5 that the condensate fraction remains to be above 60% in the performed experiment, even despite the presence of quite strong disorder.

Secondly, we examine the fluctuations of the cloud widths and their dependence on the disorder strength, since they are naturally accessible in the experiment. For all the strengths, the distributions of the widths can be well approximated by a Gaussian, see figure 7(a). The obtained distributions are compared based on the ratio of their standard deviation to the respective average value, normalized to the relative fluctuations in the clean case. With increasing disorder strength, the relative fluctuations increase approximately linearly for both transverse directions, as shown in figure 7(b). Notably, the smaller width in *x*-direction turns out to be more sensitive to the disorder variations than the larger one in *y*-direction. This is not unexpected, since the cloud is much narrower in *x*-direction and, hence, more disturbed by the laser speckle potential landscape. Note that the cloud size may vary not only due to different disorder realizations, but also due to the fluctuating particle number. Similarly as in the clean case Thomas–Fermi limit, we find that in the presence of disorder the widths are closely proportional to $N^{1/5}$. Therefore, we also present the relative fluctuations of $N^{1/5}$, given by the triangles in figure 7(b). The respective distribution is nearly constant and clearly excludes the possibility that the cloud widths fluctuate due to a change of the number of molecules.

## 5. Conclusion

Within a theory-experiment collaboration we have performed a detailed case study of the dirty boson problem by analyzing the cloud shape of a molecular BEC in a disordered trap. The geometric mean of the experimentally measured transverse cloud widths, which corresponds to the cross-section area of the cloud in the imaging plane, turned out to perfectly agree within the error bars with a theory, which is non-perturbative with respect to disorder and includes quantum fluctuations. However, the experimentally measured and theoretically calculated individual cloud widths display deviations, which we attributed to the theoretical limitations of treating the harmonic confinement within LDA. Furthermore, we also investigated the fluctuations of the measured cloud widths and showed that they do not stem from particle number fluctuations, but only depend on the disorder strength. This experimental observation goes beyond the scope of the present theory, which only considers disorder-ensemble averaged quantities, without their corresponding fluctuations. Thus, we conclude that the performed experiment on a molecular BEC in a disordered trap paves the way and asks for a more refined theory in future studies.

## Acknowledgments

We thank Benjamin Gänger and Jan Phieler for their help in initially constructing the experiment and Tobias Lausch for developing the software that runs the lab. Furthermore, we thank Antun Balaž, Piet Brouwer, and Aristeu Lima for insightful discussions. This work was supported by the Deutsche Forschungsgemeinschaft (DFG, German Research Foundation) via the Collaborative Research Center SFB/TR185 (Project No. 277625399). BN receives support from a DFG Fellowship through the Excellence Initiative by the Graduate School Materials Science in Mainz (GSC 266).

## Appendix A. Cumulant expansion method

Here, we introduce the basics of the cumulant expansion method and provide some examples that are used in our calculations. Let $X_1, X_2, \ldots, X_n$ denote random variables. Their respective moments (averages) are defined by the moment generating function

$$M(z_1, z_2, \ldots, z_n) = \langle \exp(z_1 X_1 + z_2 X_2 + \ldots + z_n X_n) \rangle$$
$$= \sum_{\nu_1=0}^{\infty} \sum_{\nu_2=0}^{\infty} \cdots \sum_{\nu_n=0}^{\infty} \frac{z_1^{\nu_1}}{\nu_1!} \frac{z_2^{\nu_2}}{\nu_2!} \cdots \frac{z_n^{\nu_n}}{\nu_n!} \langle X_1^{\nu_1} X_2^{\nu_2} \cdots X_n^{\nu_n} \rangle, \tag{A.1}$$

while the cumulants are defined by the cumulant generating function

$$K(z_1, z_2, \ldots, z_n) = \log M(z_1, z_2, \ldots, z_n)$$
$$= \sum_{\nu_1=0}^{\infty} \sum_{\nu_2=0}^{\infty} \cdots \sum_{\nu_n=0}^{\infty}{}' \frac{z_1^{\nu_1}}{\nu_1!} \frac{z_2^{\nu_2}}{\nu_2!} \cdots \frac{z_n^{\nu_n}}{\nu_n!} \langle X_1^{\nu_1}, X_2^{\nu_2}, \ldots, X_n^{\nu_n} \rangle_c, \tag{A.2}$$

where the prime sign denotes that the term with $\nu_1 = \nu_2 = \ldots = \nu_n = 0$ is excluded from the summation [75]. Moments can be expressed by cumulants of equal and lower orders and vice versa. For convenience, we note the following lowest-order examples:

$$\langle X_1 \rangle = \langle X_1 \rangle_c, \tag{A.3}$$

$$\langle X_1 X_2 \rangle = \langle X_1 \rangle \langle X_2 \rangle + \langle X_1, X_2 \rangle_c, \tag{A.4}$$

$$\langle X_1 X_2 X_3 \rangle = \langle X_1 \rangle \langle X_2 \rangle \langle X_3 \rangle + \langle X_3 \rangle \langle X_1, X_2 \rangle_c + \langle X_2 \rangle \langle X_1, X_3 \rangle_c$$
$$+ \langle X_1 \rangle \langle X_2, X_3 \rangle_c + \langle X_1, X_2, X_3 \rangle_c, \tag{A.5}$$

$$\langle X_1 X_2 X_3 X_4 \rangle = \langle X_1 \rangle \langle X_2 \rangle \langle X_3 \rangle \langle X_4 \rangle + \langle X_3 \rangle \langle X_4 \rangle \langle X_1, X_2 \rangle_c$$
$$+ \langle X_2 \rangle \langle X_4 \rangle \langle X_1, X_3 \rangle_c + \langle X_2 \rangle \langle X_3 \rangle \langle X_1, X_4 \rangle_c$$
$$+ \langle X_1 \rangle \langle X_4 \rangle \langle X_2, X_3 \rangle_c + \langle X_1, X_4 \rangle_c \langle X_2, X_3 \rangle_c$$
$$+ \langle X_1 \rangle \langle X_3 \rangle \langle X_2, X_4 \rangle_c + \langle X_1, X_3 \rangle_c \langle X_2, X_4 \rangle_c$$
$$+ \langle X_1 \rangle \langle X_2 \rangle \langle X_3, X_4 \rangle_c + \langle X_1, X_2 \rangle_c \langle X_3, X_4 \rangle_c$$
$$+ \langle X_4 \rangle \langle X_1, X_2, X_3 \rangle_c + \langle X_3 \rangle \langle X_1, X_2, X_4 \rangle_c$$
$$+ \langle X_2 \rangle \langle X_1, X_3, X_4 \rangle_c + \langle X_1 \rangle \langle X_2, X_3, X_4 \rangle_c$$
$$+ \langle X_1, X_2, X_3, X_4 \rangle_c. \tag{A.6}$$

In the Gaussian approximation the cumulants of higher than second order are neglected, so that (A.5) and (A.6) reduce to

$$\langle X_1 X_2 X_3 \rangle \approx \langle X_1 \rangle \langle X_2 \rangle \langle X_3 \rangle + \langle X_3 \rangle \langle X_1, X_2 \rangle_c + \langle X_2 \rangle \langle X_1, X_3 \rangle_c + \langle X_1 \rangle \langle X_2, X_3 \rangle_c$$
$$= \langle X_3 \rangle \langle X_1 X_2 \rangle + \langle X_2 \rangle \langle X_1 X_3 \rangle + \langle X_1 \rangle \langle X_2 X_3 \rangle - 2 \langle X_1 \rangle \langle X_2 \rangle \langle X_3 \rangle, \tag{A.7}$$

$$\langle X_1 X_2 X_3 X_4 \rangle \approx \langle X_1 \rangle \langle X_2 \rangle \langle X_3 \rangle \langle X_4 \rangle + \langle X_3 \rangle \langle X_4 \rangle \langle X_1, X_2 \rangle_c + \langle X_2 \rangle \langle X_4 \rangle \langle X_1, X_3 \rangle_c$$
$$+ \langle X_2 \rangle \langle X_3 \rangle \langle X_1, X_4 \rangle_c + \langle X_1 \rangle \langle X_4 \rangle \langle X_2, X_3 \rangle_c + \langle X_1, X_4 \rangle_c \langle X_2, X_3 \rangle_c$$
$$+ \langle X_1 \rangle \langle X_3 \rangle \langle X_2, X_4 \rangle_c + \langle X_1, X_3 \rangle_c \langle X_2, X_4 \rangle_c + \langle X_1 \rangle \langle X_2 \rangle \langle X_3, X_4 \rangle_c$$
$$+ \langle X_1, X_2 \rangle_c \langle X_3, X_4 \rangle_c = \langle X_1 X_3 \rangle \langle X_2 X_4 \rangle + \langle X_1 X_2 \rangle \langle X_3 X_4 \rangle$$
$$+ \langle X_2 X_3 \rangle \langle X_1 X_4 \rangle - 2 \langle X_1 \rangle \langle X_2 \rangle \langle X_3 \rangle \langle X_4 \rangle, \tag{A.8}$$

where (A.3) and (A.4) were invoked to get the final results. As a special case, we obtain the approximations of the cubic and quartic moments of a random variable

$$\langle X_1^3 \rangle \approx 3 \langle X_1 \rangle \langle X_1^2 \rangle - 2 \langle X_1 \rangle^3, \tag{A.9}$$

$$\langle X_1^4 \rangle \approx 3 \langle X_1^2 \rangle^2 - 2 \langle X_1 \rangle^4. \tag{A.10}$$

For the sake of generality, we note that within the Gaussian approximation the following recursive relation holds

$$\langle X_1^k \rangle = \langle X_1 \rangle \langle X_1^{k-1} \rangle + (k-1)(\langle X_1^2 \rangle - \langle X_1 \rangle^2)\langle X_1^{k-2} \rangle, \quad k \geqslant 2, \tag{A.11}$$

where $\langle X_1^0 \rangle = 1$.

## Appendix B. Non-perturbative treatment of the disorder

Here, we briefly go through the main steps of our non-perturbative account of the disorder potential in a theory that also contains the quantum fluctuations corrections. Of course, the same procedure is applicable to a simple Gross–Pitaevskii equation [83]. Under the approximations mentioned in the main text, (8) becomes

$$-\frac{\hbar^2}{2m}\nabla_\perp^2 \varphi(\mathbf{r}_\perp) + U\varphi(\mathbf{r}_\perp) + (u(\mathbf{r}_\perp) - \langle u \rangle)\langle \psi \rangle + (u(\mathbf{r}_\perp)\varphi(\mathbf{r}_\perp) - \langle u(\mathbf{r}_\perp)\varphi(\mathbf{r}_\perp)\rangle) - \mu\varphi(\mathbf{r}_\perp)$$
$$+ g\left\{3\langle\psi\rangle^2\varphi(\mathbf{r}_\perp) + 3\langle\psi\rangle(\varphi(\mathbf{r}_\perp)^2 - \langle\varphi^2\rangle) + \varphi(\mathbf{r}_\perp)^3 + \frac{40}{3\sqrt{\pi}}a_s^{3/2}[4\langle\psi\rangle^3\varphi(\mathbf{r}_\perp)\right.$$
$$\left. + 6\langle\psi\rangle^2(\varphi(\mathbf{r}_\perp)^2 - \langle\varphi^2\rangle) + 4\langle\psi\rangle\varphi(\mathbf{r}_\perp)^3 + (\varphi(\mathbf{r}_\perp)^4 - 3\langle\varphi^2\rangle^2)]\right\} = 0, \tag{B.1}$$

where now there is only a transverse spatial dependence due to the assumed two-dimensional character of the disorder and due to the LDA. Hence, we are here effectively solving the problem of a homogeneous two-dimensional disordered Bose gas. An extension to three dimensions is straightforward. Note that the cumulant expansion has also been used. Multiplying the previous equation with $u(\mathbf{r}'_\perp)$ and averaging produces

$$-\frac{\hbar^2}{2m}\nabla_\perp^2 \langle u(\mathbf{r}'_\perp)\varphi(\mathbf{r}_\perp)\rangle + U\langle u(\mathbf{r}'_\perp)\varphi(\mathbf{r}_\perp)\rangle + (\langle u(\mathbf{r}'_\perp)u(\mathbf{r}_\perp)\rangle - \langle u\rangle^2)\langle\psi\rangle$$
$$+ (\langle u(\mathbf{r}'_\perp)u(\mathbf{r}_\perp)\varphi(\mathbf{r}_\perp)\rangle - \langle u\rangle\langle u(\mathbf{r}_\perp)\varphi(\mathbf{r}_\perp)\rangle) - \mu\langle u(\mathbf{r}'_\perp)\varphi(\mathbf{r}_\perp)\rangle$$
$$+ g\left\{3\langle\psi\rangle^2\langle u(\mathbf{r}'_\perp)\varphi(\mathbf{r}_\perp)\rangle + 3\langle\psi\rangle(\langle u(\mathbf{r}'_\perp)\varphi(\mathbf{r}_\perp)^2\rangle - \langle u\rangle\langle\varphi^2\rangle) + \langle u(\mathbf{r}'_\perp)\varphi(\mathbf{r}_\perp)^3\rangle\right.$$
$$+ \frac{40}{3\sqrt{\pi}}a_s^{3/2}[4\langle\psi\rangle^3\langle u(\mathbf{r}'_\perp)\varphi(\mathbf{r}_\perp)\rangle + 6\langle\psi\rangle^2(\langle u(\mathbf{r}'_\perp)\varphi(\mathbf{r}_\perp)^2\rangle - \langle u\rangle\langle\varphi^2\rangle)$$
$$\left. + 4\langle\psi\rangle\langle u(\mathbf{r}'_\perp)\varphi(\mathbf{r}_\perp)^3\rangle + (\langle u(\mathbf{r}'_\perp)\varphi(\mathbf{r}_\perp)^4\rangle - 3\langle u\rangle\langle\varphi^2\rangle^2)]\right\} = 0, \tag{B.2}$$

which reduces after the cumulant expansion to

$$-\frac{\hbar^2}{2m}\nabla_\perp^2 \langle u(\mathbf{r}'_\perp), \varphi(\mathbf{r}_\perp)\rangle_c + (V + V_Q)\langle u(\mathbf{r}'_\perp), \varphi(\mathbf{r}_\perp)\rangle_c + \langle u(\mathbf{r}'_\perp), u(\mathbf{r}_\perp)\rangle_c \langle\psi\rangle = 0, \tag{B.3}$$

where $V = U + \langle u \rangle - \mu + 3g\langle\psi^2\rangle$ and $V_Q = \frac{160g}{3\sqrt{\pi}}a_s^{3/2}\langle\psi\rangle(3\langle\psi^2\rangle - 2\langle\psi\rangle^2)$ have been introduced in the main text. Similarly, multiplying (B.1) with $\varphi(\mathbf{r}'_\perp)$ and averaging gives

$$-\frac{\hbar^2}{2m}\nabla_\perp^2 \langle\varphi(\mathbf{r}'_\perp)\varphi(\mathbf{r}_\perp)\rangle + U\langle\varphi(\mathbf{r}'_\perp)\varphi(\mathbf{r}_\perp)\rangle + \langle\varphi(\mathbf{r}'_\perp)u(\mathbf{r}_\perp)\rangle\langle\psi\rangle + \langle\varphi(\mathbf{r}'_\perp)u(\mathbf{r}_\perp)\varphi(\mathbf{r}_\perp)\rangle$$
$$- \mu\langle\varphi(\mathbf{r}'_\perp)\varphi(\mathbf{r}_\perp)\rangle + g\left\{3\langle\psi\rangle^2\langle\varphi(\mathbf{r}'_\perp)\varphi(\mathbf{r}_\perp)\rangle + 3\langle\psi\rangle\langle\varphi(\mathbf{r}'_\perp)\varphi(\mathbf{r}_\perp)^2\rangle\right.$$
$$+ \langle\varphi(\mathbf{r}'_\perp)\varphi(\mathbf{r}_\perp)^3\rangle + \frac{40}{3\sqrt{\pi}}a_s^{3/2}[4\langle\psi\rangle^3\langle\varphi(\mathbf{r}'_\perp)\varphi(\mathbf{r}_\perp)\rangle + 6\langle\psi\rangle^2\langle\varphi(\mathbf{r}'_\perp)\varphi(\mathbf{r}_\perp)^2\rangle$$
$$\left. + 4\langle\psi\rangle\langle\varphi(\mathbf{r}'_\perp)\varphi(\mathbf{r}_\perp)^3\rangle + \langle\varphi(\mathbf{r}'_\perp)\varphi(\mathbf{r}_\perp)^4\rangle]\right\} = 0, \tag{B.4}$$

which simplifies after the cumulant expansion to

$$-\frac{\hbar^2}{2m}\nabla_\perp^2 \langle\varphi(\mathbf{r}'_\perp), \varphi(\mathbf{r}_\perp)\rangle_c + (V + V_Q)\langle\varphi(\mathbf{r}'_\perp), \varphi(\mathbf{r}_\perp)\rangle_c + \langle u(\mathbf{r}'_\perp), \varphi(\mathbf{r}_\perp)\rangle_c \langle\psi\rangle = 0. \tag{B.5}$$

Note that in the last term we replaced $\langle u(\mathbf{r}_\perp), \varphi(\mathbf{r}'_\perp)\rangle_c$ with $\langle u(\mathbf{r}'_\perp), \varphi(\mathbf{r}_\perp)\rangle_c$, which should be exact in the homogeneous case. The obtained two coupled equations (B.3) and (B.5) can then be rewritten as

$$-\frac{\hbar^2}{2m}\nabla_\perp^2 \langle u, \varphi\rangle_c(\mathbf{r}'_\perp - \mathbf{r}_\perp) + (V + V_Q)\langle u, \varphi\rangle_c(\mathbf{r}'_\perp - \mathbf{r}_\perp) + \langle u, u\rangle_c(\mathbf{r}'_\perp - \mathbf{r}_\perp)\langle\psi\rangle = 0, \tag{B.6}$$

$$-\frac{\hbar^2}{2m}\nabla_\perp^2 \langle\varphi, \varphi\rangle_c(\mathbf{r}'_\perp - \mathbf{r}_\perp) + (V + V_Q)\langle\varphi, \varphi\rangle_c(\mathbf{r}'_\perp - \mathbf{r}_\perp) + \langle u, \varphi\rangle_c(\mathbf{r}'_\perp - \mathbf{r}_\perp)\langle\psi\rangle = 0, \tag{B.7}$$

due to spatial homogeneity after the disorder averaging. In the transverse $\mathbf{k}_\perp$-plane one finds

$$(\varepsilon_{\mathbf{k}_\perp} + V + V_Q)\langle u, \varphi\rangle_c(\mathbf{k}_\perp) + \langle u, u\rangle_c(\mathbf{k}_\perp)\langle\psi\rangle = 0, \tag{B.8}$$

$$(\varepsilon_{\mathbf{k}_\perp} + V + V_Q)\langle\varphi, \varphi\rangle_c(\mathbf{k}_\perp) + \langle u, \varphi\rangle_c(\mathbf{k}_\perp)\langle\psi\rangle = 0, \tag{B.9}$$

where $\varepsilon_{\mathbf{k}_\perp} = \hbar^2 \mathbf{k}_\perp^2/(2m) = \hbar^2(k_x^2 + k_y^2)/(2m)$ denotes the transverse dispersion, so that

$$\langle u, \varphi \rangle_c (\mathbf{k}_\perp) = -\frac{\langle \psi \rangle \langle u, u \rangle_c (\mathbf{k}_\perp)}{\varepsilon_{\mathbf{k}_\perp} + V + V_Q}, \tag{B.10}$$

$$\langle \varphi, \varphi \rangle_c (\mathbf{k}_\perp) = \frac{\langle \psi \rangle^2 \langle u, u \rangle_c (\mathbf{k}_\perp)}{(\varepsilon_{\mathbf{k}_\perp} + V + V_Q)^2}. \tag{B.11}$$

Integration over the entire $\mathbf{k}_\perp$-plane of the previous equations finally yields

$$\langle u(\mathbf{r}_\perp), \varphi(\mathbf{r}_\perp) \rangle_c = -\langle \psi \rangle \int_{\mathbb{R}^2} \frac{d^2 \mathbf{k}_\perp}{(2\pi)^2} \frac{\langle u, u \rangle_c (\mathbf{k}_\perp)}{\varepsilon_{\mathbf{k}_\perp} + V + V_Q}, \tag{B.12}$$

$$\langle \varphi(\mathbf{r}_\perp), \varphi(\mathbf{r}_\perp) \rangle_c = \langle \psi \rangle^2 \int_{\mathbb{R}^2} \frac{d^2 \mathbf{k}_\perp}{(2\pi)^2} \frac{\langle u, u \rangle_c (\mathbf{k}_\perp)}{(\varepsilon_{\mathbf{k}_\perp} + V + V_Q)^2}. \tag{B.13}$$

The obtained cumulants are homogeneous across the entire $\mathbf{r}_\perp$-space and depend parametrically on $\langle \psi \rangle$, $\langle \psi^2 \rangle$ and $U$. In a trapped system according to the LDA, we get

$$\langle u(\mathbf{r}), \varphi(\mathbf{r}) \rangle_c = -\langle \psi(\mathbf{r}) \rangle \int_{\mathbb{R}^2} \frac{d^2 \mathbf{k}_\perp}{(2\pi)^2} \frac{\langle u, u \rangle_c (\mathbf{k}_\perp)}{\varepsilon_{\mathbf{k}_\perp} + V(\mathbf{r}) + V_Q(\mathbf{r})}, \tag{B.14}$$

$$\langle \varphi(\mathbf{r}), \varphi(\mathbf{r}) \rangle_c = \langle \psi(\mathbf{r}) \rangle^2 \int_{\mathbb{R}^2} \frac{d^2 \mathbf{k}_\perp}{(2\pi)^2} \frac{\langle u, u \rangle_c (\mathbf{k}_\perp)}{(\varepsilon_{\mathbf{k}_\perp} + V(\mathbf{r}) + V_Q(\mathbf{r}))^2}, \tag{B.15}$$

by restoring the explicit spatial dependence of $\langle \psi(\mathbf{r}) \rangle$, $\langle \psi^2(\mathbf{r}) \rangle$ and $U(\mathbf{r})$. The last two equations correspond to (11) and (12) of the main text.

## ORCID iDs

Benjamin Nagler  https://orcid.org/0000-0002-6961-0734
Milan Radonjić  https://orcid.org/0000-0002-2972-2969
Artur Widera  https://orcid.org/0000-0002-0338-9969